# Quantum Monte Carlo study of low dimensional magnetic system

A.K. Murtazaev, M.A. Magomedov

*Institute of Physics, Dagestan Scientific Center, Russian Academy of Sciences, Yaragskogo Str. 94, Makhachkala 367003, Russia*

*e-mail: magomedov_ma@mail.ru*

**Abstract:** Monte Carlo simulations are performed for the $S = ½$ XY and ferro- and antiferromagnetic Heisenberg model in two dimensions using the loop algorithm. Thermodynamic properties of all these models are investigated in wide temperature range. The energy, specific heat, susceptibility and other parameters are given as function of temperature and the Trotter number. The comparison of calculated thermodynamic quantities with theoretical and with experimental data are given. It is shown that our results are in good agreement with them.

## Introduction

The investigation of thermodynamic properties for spin systems in view of their quantum character is one of the important problems of modern condensed matter physics. However, unlike classical systems, one meets serious difficulties at investigation of quantum spin systems. In present years the considerable attention is paid to study these systems.

One of the important problems is a nature of a ground state of two-dimensional antiferromagnetic Heisenberg model. The numerical calculations and analysis of experimental data testify to existence of the long-range order at zero temperature in Heisenberg antiferromagnets. However the majority of exact calculations are gained for the system with small size of lattices. The evolutions of computation techniques allow using great lattice size that gives decreasing of the long-range antiferromagnetic order as $\sim 1/N$.

One of the powerful techniques to investigate the quantum spin systems is a quantum version of Monte Carlo method, so-called quantum Monte Carlo method (QMC) [2-5]. The method directly benefits from a progress of microprocessors. It is also suitable for the parallel computing. Developments of various Monte Carlo algorithms consist of an important part of the statistical physics in these decades.

However, despite of considerable efforts, the opportunity of effective application of QMC method to study the properties of quantum systems, in particular, lattice models with arbitrary value of a spin in a wide interval of change of various physical parameters, still remains problematic.

One of effective research techniques of such systems is a method based on the decomposition Suzuki-Trotter formula [1]:

$$\exp\left(\sum_{i=1}^{p} A_i\right) = \lim_{m \to \infty} \left(e^{A_1/m} \times e^{A_2/m} \times \cdots e^{A_p/m}\right)^m, \quad (1)$$

where $m$ is the integer positive number (Trotter number).

The partition function is estimated by using this formula, and finite-temperature properties are calculated. This method has been successfully applied to various problems. The Suzuki-Trotter formula transforms any $d$-dimensional quantum system into a $(d+1)$-dimensional classical system of a general type.

The additional dimension is called the Trotter direction and its length is denoted by the Trotter number $m$. The original quantum system is recovered in the limit of the infinite Trotter number.

The first update algorithm in QMC is the Worldline algorithm with local flip of spins [3-5]. A Worldline, which is an outcome of the local spin conservation, is moved locally in the $(d+1)$-dimensional lattice. An acceptance ratio of this flip becomes worse and the Monte Carlo dynamics freezes as the Trotter number increases to approach the original quantum spin system. This is purely a technical problem not from a physical origin as the critical slowing down.

Recently, the loop algorithm of QMC method with nonlocal flip is developed in order to solve this problem. It is also possible to take the infinite Trotter number limit beforehand. A nonlocal loop is defined in the $(d+1)$-dimensional lattice for the update. All spins on the loop are simultaneously flipped. This loop corresponds to the correlated spins in this $(d+1)$-dimensional lattice. Therefore, the algorithm drastically reduces the correlation time. The loop algorithms, or generally cluster algorithms, are a trend of recent developments in Monte Carlo methods. The single-cluster variant of loop algorithm in details presented in [2].

**Models and Method**

Last years many investigations are devoted to theoretical and experimental examination of two-dimensional XY ($2d$ XY), ferromagnetic ($2d$ FH) and antiferromagnetic ($2d$ AFH) Heisenberg model with a spin $S = ½$.

We studied quantum XY, and ferro- and antiferromagnetic Heisenberg model on a square lattice with a spin $S = ½$ using highly effective loop algorithm [2], based on the Suzuki-Trotter approach. The investigation were carried out for systems with linear size $L \times L \times 4m$ with $L = 32$ and $m = 8, 16, 32$ and $64$. Thus, the number of spins in system was respectively equal to $N = 32768, 65536, 131072$ and $262144$. At the system the periodic boundary conditions were applied in real and Trotter directions.

We consider isotropic XY and ferro- and antiferromagnetic Heisenberg model with a spin $S = ½$ and with interaction between the nearest neighbors on a square lattice. The Hamiltonian of this model can be presented in the following view [3]:

$$H = -J\sum_{i,j}\left(S_i^x S_j^x + S_i^y S_j^y + \Delta S_i^z S_j^z\right), \qquad (2)$$

where $J > 0$, and $\Delta = 0$, 1 and -1 for XY, ferromagnetic and antiferromagnetic Heisenberg model with an appropriate unitary transformation respectively [3-5]. We treat the intermediate models ($\Delta = \pm 0.5$) as well.

The basic thermodynamic quantities (such as energy $E$, specific heat $C$, magnetization $M$, susceptibility $\chi$) can be calculated by means of fluctuation relations:

$$E^{(m)} = \left\langle F^{(m)} \right\rangle, \qquad (3)$$

$$C^{(m)}T^2 = \left\langle F^{(m)^2} - G^{(m)} \right\rangle - \left\langle F^{(m)} \right\rangle^2, \qquad (4)$$

$$M_k^{(m)} = \frac{1}{4m}\left[\sum_i \sum_j S_{i,j}\right]_k, \qquad (5)$$

$$\chi^{(m)} = \beta\left\langle M^{(m)^2} \right\rangle - \left\langle M^{(m)} \right\rangle^2, \qquad (6)$$

$$N_C^{(m)} = \frac{1}{N}\left\langle N_C \right\rangle, \qquad (7)$$

where $T$ is temperature, $\beta$ is inverse temperature, $M_k^{(m)}$ is a magnetic moment of system in $k$-th a state, $N$ is number of spins in system, $N_C^{(m)}$ is mean size of a cluster. The values $F^{(m)}$, $G^{(m)}$ are given, for example, in [4, 5].

Also the loop algorithm allows to use for evaluation of thermodynamic quantities the so-called improved estimators, which are characterized by prompt convergence and high accuracy in comparison with the traditional formulas. Especially simple estimators can be derived for magnetic susceptibility. For example, the uniform magnetic susceptibility at vanishing magnetic field can be expressed as [2]:

$$\chi_{impr}^{(m)} = 2dN\beta\left\langle \frac{M_C^2}{|N_C|} \right\rangle, \qquad (8)$$

where $M_C$ is a magnetization of a cluster, $|N_C|$ is the size of a cluster.

It is necessary to note, that average quantities computed by fluctuation relations (3-8) will correspond to the precise values at $m \to \infty$. Especially strongly it is exhibited at low temperatures and for small $m$. Thus, for definition of precise value of thermodynamic quantities was used linear or square-law extrapolation [3, 5]. Treating the Trotter size *m* up to 64, we follow the extrapolation procedure to get the $m \to \infty$ value of thermodynamic quantities. For example, for energy it is possible to show:

$$E(m) = E(\infty) + a/m^2 + b/m^4 . \qquad (9)$$

## Results

The temperature dependences of energy *E* for antiferromagnetic Heisenberg model at various values of Trotter number *m* and also after extrapolation ($m \to \infty$) by using formula (9) are given in figure 1. We measured the energy and temperature in units of *J*. For extrapolation of data we used three different values of the Trotter number: $m = 16$, $m = 32$ and $m = 64$. The data from modified spin-wave theory, which developed by Takahashi, are also given. The modified spin-wave theory is expected to be valid in the low-temperature regime. The agreement with our calculation is rather good.

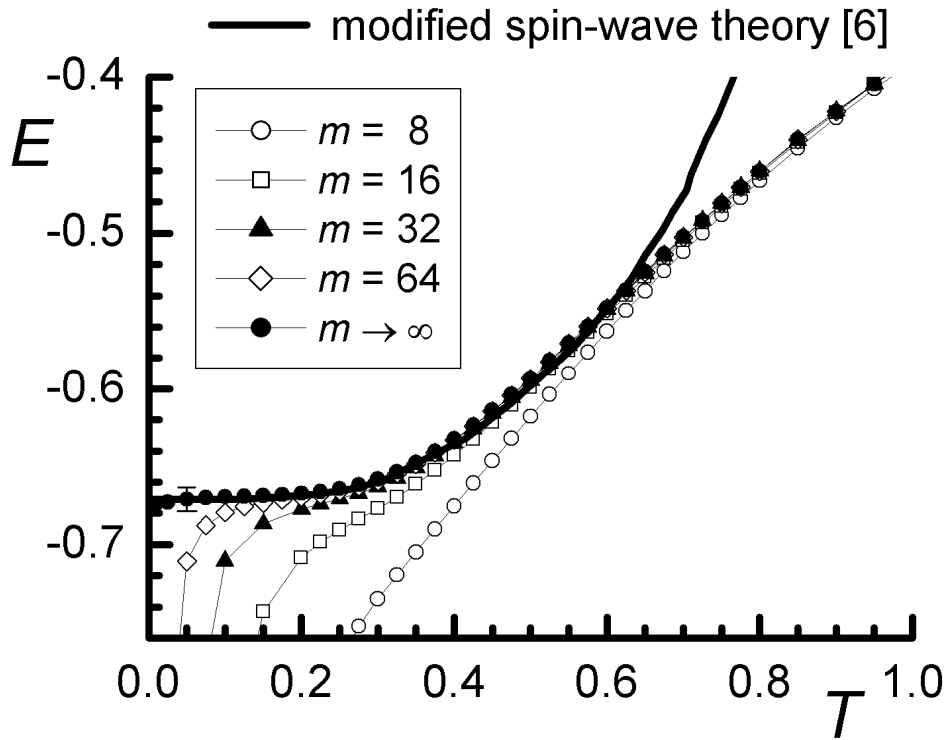

Fig. 1. Temperature dependence of energy *E* for antiferromagnetic Heisenberg model at various *m*.

The temperature dependences of energy *E* for various models in wide temperature range are given in figure 2. The data from modified spin-wave theory and high-temperature series expansion are also given. The modified spin-wave theory is good in the low-temperature regime. The high-temperature series expansion works well in the high-temperature limit. The Monte-Carlo method is excellent work as well as in low and high temperature regime.

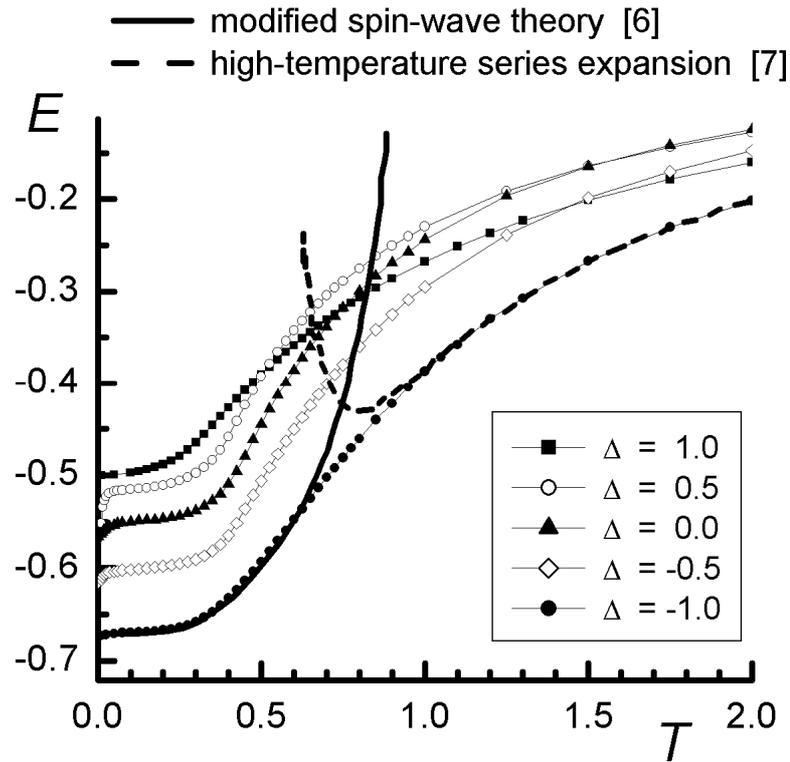

Fig. 2. Temperature dependence of energy $E$ for various models (various $\Delta$).

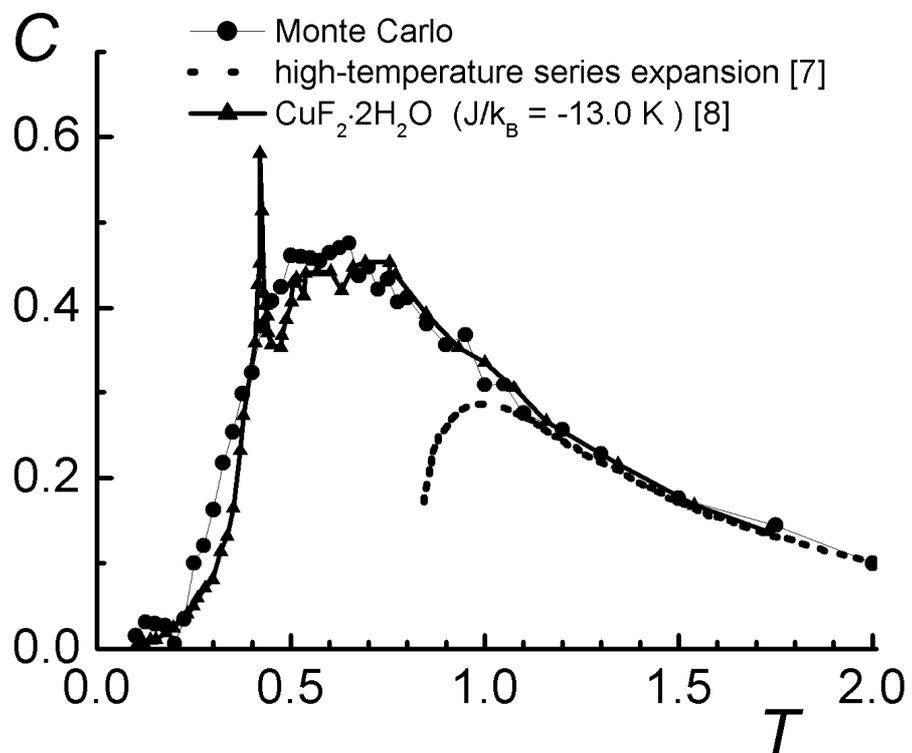

Fig. 3. Temperature dependence of specific heat for antiferromagnetic Heisenberg model.

The specific heat data for antiferromagnetic Heisenberg model are shown in figure 3. The data from high-temperature series expansion and experimental data for one of typical quasi two-dimensional magnetic materials are also presented. The agreement with high-temperature series expansion, in the range $T \geq 1.1$, and experimental data are excellent.

In figure 4 are shown the temperature dependences of susceptibility $\chi$ for antiferromagnetic Heisenberg model. Presented our data are calculated by using formula (8) and extrapolated to ($m \to \infty$) by formula (9). Also the data from modified spin-wave theory, high-temperature series expansion and experimental data are given.

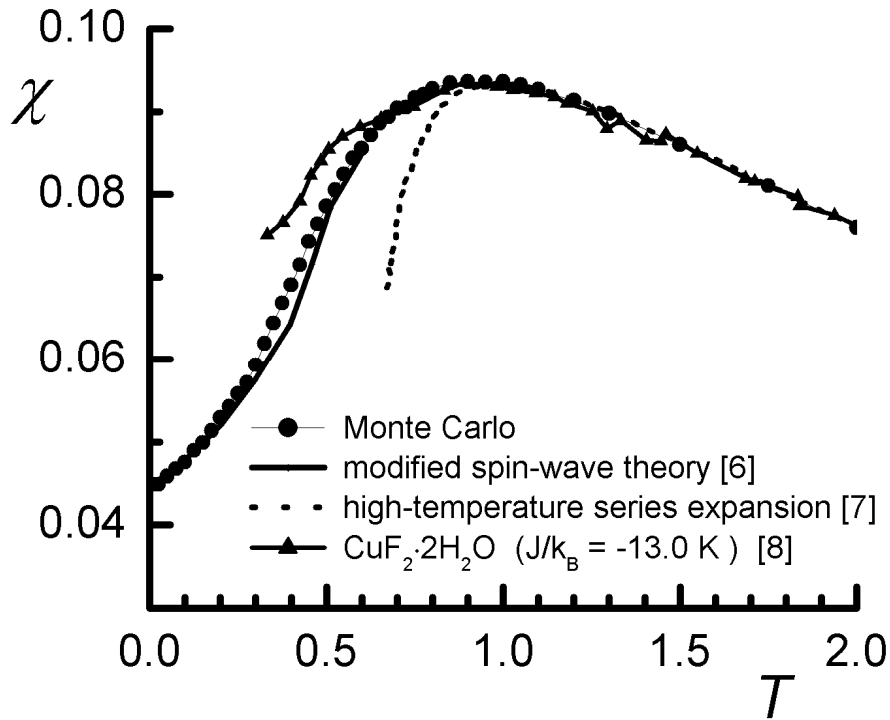

Fig. 4. Temperature dependence of susceptibility $\chi$ for antiferromagnetic Heisenberg model.

**Conclusions**

Extensive Monte Carlo simulations are performed for the various quantum spin $S = \frac{1}{2}$ two-dimensional models (XY, ferro- and antiferromagnetic Heisenberg model) using the high-accuracy loop algorithm. Thermodynamic properties of these models are investigated in wide temperature range. The energy $E$, specific heat $C$, susceptibility $\chi$ are calculated.

The carried out comparison of calculated thermodynamic quantities with theoretical and with experimental result is given for some quasi two-dimensional magnetic systems. It is shown that results got on base of the Quantum Monte-Carlo methods, are found to be in good agreement with theoretical and experimental data and well described the thermodynamic properties of low-dimensional magnetic systems.

## Acknowledgements

This work is supported by the Russian Foundation of Basic Research (project № 04-02-16487), FCP "Integration" (№ I0228), by grant of scientific school (№ SS–2253.2003.2) and by the program of "Russian Science Support Foundation" (A.K. Murtazaev).